**AI as Teammate: Rethinking Task Distribution in Medical Training**

Fendi Tsim[1*†], Alina Gutoreva[2*], Anthony Weiss[3], Nicole Dubosh[3]

## Abstract

Integrating Artificial Intelligence (AI), particularly generative AI, into medical training has prompted widespread cognitive and behavioral concerns about learner over-reliance, misuse, and the erosion of foundational clinical competencies. In this paper, we propose a conceptual reframing at the decision level: the problem is not misuse but *misclassification*: a mechanistic failure of real-time metacognitive evaluation in selecting a subzone-inappropriate AI interaction mode. Drawing on "SCAN" (Substitute, Complement, Aid, Non-Negotiable) — a human-centric decision-making framework for Generative AI task allocation based on Vygotsky's Zone of Proximal Development and Metacognition, we advance the emerging social-constructivist conversation around AI in medical education, by offering a testable account of the role of AI in clinical reasoning development. This framework yields testable predictions for how misclassification can be detected, mitigated, and more importantly, prevented in the clinical learning environment. As for clinical reasoning development, we illustrate how trajectories of skill acquisition (e.g., upskilling) and failure (e.g., the triad of skill failure (de-skilling, never-skilling, and mis-skilling)) operate at the task level of an individual that fixed-phase, cohort-wide treatments fail to account for. We further identify passive engagement within correctly classified AI-scaffolded tasks as a particularly insidious and detection-resistant pathway to mis-skilling — one that requires subzone re-identification from AI assistance to expert assistance, with human experts serving as epistemic auditors. The paper operationalizes SCAN for the clinical curriculum design, supervision, and assessment, and opens an empirical research agenda grounded in cognitive science (e.g. learning, decision-making). The paradigm shift from misuse to misclassification is not semantic: it offers, we believe, a clear perspective for educators to change what to look for, what to assess, and what to intervene on.

---

[1] Behavioral AI Institute, London, UK
[*] These authors contributed equally to this work.
[†] Correspondence to: Fendi Tsim (fenditsim@gmail.com)
[2] College of Human Sciences and Education, KIMEP University, 050010, Abay Avenue 2, Almaty, Kazakhstan
[3] Harvard Medical School, 25 Shattuck Street Boston, MA 02115, USA

## 1. Introduction

It is commonly known that Artificial Intelligence (AI) has entered the medical training environment faster than the conceptual vocabulary needed to evaluate its effects (Feng & Shen, 2023; van de Ridder et al., 2023). This speed, of course, carries severe consequences. A recent survey found that while 97.1% of graduate medical trainees report no formal AI instruction, 85.5% report already using AI tools for clinical decision support, academic writing and research (Benton et al., 2026). Notably, their survey also found that over a third of high-familiarity trainees report low confidence, suggesting exposure alone does not build the capacity to evaluate the AI output critically. Apart from that, educators observe learners delegating tasks to generative AI (henceforth GenAI) that they should be performing themselves (Izquierdo-Condoy et al., 2025). A second-year resident, albeit the AI-generated note is "flawless" in structure and detail, struggles to articulate the reasoning behind it when asked directly (Preiksaitis, 2026). Researchers document automation bias (Khera et al., 2023; Nguyen, 2024), over-reliance (Janumpally et al., 2025), and skill erosion (Macnamara et al., 2024) — consistent with AI adoption in medical education that, notably, remains informal, and more importantly, without structure oversight (Hernández Rincón et al., 2025; Stamos & Dharod, 2026).

The field identified, and thus has begun to address, this problem at the *structural* level: institutions respond with policies ranging from outright prohibition to enthusiastic integration (Kung et al., 2023; Mbakwe et al., 2023). Quillen College of Medicine at East Tennessee State University has developed a five-step AI integration process linking policy to access decisions (Rueff et al., 2025). Building on an earlier pilot demonstrating feasibility, Geisel School of Medicine at Dartmouth has announced a longitudinal AI curriculum designed to produce digital health leaders across all four years of training (Dartmouth Health & Geisel School of Medicine Giving, 2024). Mass General Brigham has moved toward health-system governance frameworks for AI use in clinical practice (Saenz et al., 2024). The Association of American Medical Colleges has published seven guiding principles, and a policy checklist for AI use in medical education (Association of American Medical Colleges, 2025). In short, the movement from Quillen (policy and access) to Geisel (curriculum) to Mass General Brigham (health-system governance) to the AAMC principles represents institutional ambition and significant progress.

Yet each of these responses, however current, shares a common analysis at the cohort-wide level, rather than at the individual level, when a learner faces a task. Unbeknownst to many educators, though, learners are making moment-by-moment decisions about AI use that existing cohort-level policies do not, and to a large extent, are not designed to, govern. What, then, governs this decision that a learner makes when opening an AI tool before a clinical task? With the rise of generative AI tools such as large language models, a growing body of empirical research raises concerns about learner's over-reliance on, and misuse of, these tools in the clinical training (hereafter, we use "AI" to refer to GenAI specifically, unless otherwise noted). AI's fluent, judgment-simulating output invites deference in ways fixed-output tools do not. Recent experimental investigations attempt at a single learner-AI interaction. Medical students exposed to misleading AI-generated explanations showed diagnostic accuracy falling below a no-explanation baseline, with confidence no longer tracking correctness (Teng et al., 2026). Explanation style alone was found to shift the trade-off between medical students' cognitive load and confidence calibration during diagnostic learning (Tang et al., 2026). This dependence on AI for routine cognitive tasks may, as a result, undermine the development of foundational clinical reasoning skills (Boscardin et al., 2024; Blanco et al., 2025; Gordon et al., 2024; Izquierdo-Condoy et al., 2025; Lucas et al., 2024; Pham et al., 2025). While these concerns are substantiated and important (American Medical Association, 2026), we note that, to the best of our knowledge, there is not a theoretical account of what, at a single task-level interaction, makes a given AI engagement developmentally *harmful*, and more importantly, what would make it developmentally *beneficial*.

In this paper, we begin by looking at the developmental trajectory through a social constructivist lens in clinical training (Haddock et al., 2023). Knowledge is co-constructed through collaborative engagement with those with greater expertise, including teachers, peers, and recently, generative AI systems in the contemporary training environment (Cho et al., 2024; Jacobs et al., 2025; Tran et al., 2025). We propose "SCAN" — a human-centric, systematic framework for effective task allocation between a learner, human experts, and AI (Tsim & Gutoreva, 2026). Based on the Vygotsky (1978)'s Zone of Proximal Development (ZPD) and metacognition (Bergamaschi Ganapini et al., 2025, Flavell, 1979, Nelson and Narens, 1990), SCAN organizes clinical tasks along four subzones: Substitute, Complement, Aid, and Non-Negotiable. As we shall see, the taxonomic structure it provides offers a clear understanding of how a learner's overreliance and misuse of AI issues occur, and more importantly, directions for solution in terms of curriculum design, supervision and assessment.

SCAN, as an extension of Vygotsky's ZPD in the age of AI with metacognition, differentiates two developmental trajectories through its four subzones. The first one is human-human learning, or known as "Internalization Trajectory" following Vygotsky's (1978) original account, in which a human expert remains the indispensable scaffold, and such scaffold fades gradually as the learner's independence grows over time. The other one is human-AI learning, denoted as “Symbiosis Trajectory” (Tsim & Gutoreva, 2026), happens when AI itself becomes part of the “persistent” scaffold, and the learner's growing independence is expressed as a deliberate, self-directed engagement with AI over time (Gutoreva et al., 2026; Tran et al., 2025). These two developmental trajectories are, we conjecture, *complementary*: a complete clinical formation requires both for learner’s clinical reasoning development in the long run (Kassab et al., 2025; Stamos & Dharod, 2026).

With SCAN, we conjecture that what the literature describes as learner’s AI over-reliance or misuse can be best understood as *misclassification* of AI use: the assignment of a subzone-inappropriate interaction mode to a clinical task, resulting in the elimination of the cognitive engagement through which clinical reasoning develops. Existing framings such as over-reliance and misuse collapse learner’s inappropriate AI use into a single behavioral pattern, without specifying its underlying psychological mechanisms. Misclassification, in contrast, treats it as a discrete failure of real-time metacognitive evaluation — one that provides distinct and falsifiable predictions for when, and how, clinical reasoning development is disrupted. We address misclassification as a *socio-cognitive* problem in clinical training by providing a socio-cognitive solution for learners, moment-by-moment, to have an effective task allocation between themselves and AI for clinical reasoning development.

Lastly, this paper examines skill acquisition and failure in the clinical reasoning development. Upskilling, according to SCAN, occurs when (1) a learner identifies, and thus engages, a task at the appropriate subzone; (2) learner’s task-specific knowledge accumulates; and (3) the task migrates toward subzones with greater independence over time. However, when this series of processes is disrupted, it produces one of the three skill failure modes (Deskilling, Never-skilling, and Mis-skilling) or known as "The Triad of Skill Failure" (Abdulnour et al., 2025; Berzin & Topol, 2025; Ke et al., 2026; Keren, Desai & West, 2026). Deskilling is a regression through disuse (Budzyń et al., 2025). Never-skilling is a failure to acquire a competency that should have formed in the first place (Ke et al., 2026). Mis-skilling is the acquisition of a flawed competency that presents as genuine learning (Teng et al., 2026). Unlike prior treatments of this triad, which suggest resolving them at the fixed, cohort-wide level, we argue each failure mode is best understood at the task level for a learner. As we shall see, each failure pathway maps onto a well-established cognitive mechanism in behavioural science. This grounds the triad, in mechanistic terms, that generates falsifiable predictions about which instances of AI engagement are developmentally harmful, and more importantly, which are not.

Our work provides three contributions that advance the field toward a testable, mechanistic pedagogy. First, this paper advances the emerging social-constructivist conversation around AI in medical education through SCAN's two complementary, developmental trajectories. It is, we propose, a necessary condition for a complete clinical reasoning formation in the age of AI. Second, we differentiate misclassification from existing framings of AI misuse as a specific, mechanistic failure of real-time metacognitive evaluation. This reframing converts a descriptive concern into a falsifiable construct with operational pathways for curriculum design, clinical supervision, and assessment. Lastly, this paper offers a testable, task-level account of skill acquisition and failure for a learner with each pathway based on a specific, well-established cognitive mechanism. These theory-driven pathways generate falsifiable predictions that could, we argue, differentiate harmful AI engagements from beneficial ones.

## 2. Theoretical Foundation

An emerging body of literature examines the use of generative AI in the clinical training via social constructivism approach (Cho et al., 2024; Jacobs et al., 2025; Tran et al., 2025). The common argument, in this view, is whether generative AI can offer great expertise by assisting learners learning as a scaffold. In human-human learning, “more knowledgeable others” (MKO) are crucial in Vygotsky (1978)’s work of ZPD—a well-known framework in illustrating an individual’s learning and development via collaborative and peer learning, with human experts such as teachers and peers who are more experienced, and thus offer greater expertise as assistance. Tsim and Gutoreva (2026) proposed a framework called “SCAN”—an extension of Vygotsky’s ZPD in the age of AI by considering generative AI as another “more knowledgeable other” apart from human experts, together with metacognition (one’s awareness about one’s own thinking).

SCAN is a human-centric, decision-making framework for systematic task allocation between a learner, human experts, and AI. It was developed to address the absence of a principled, task-level vocabulary for AI integration in cognitively demanding professional domains. The framework organizes tasks along a continuum defined by two dimensions: the degree to which AI can appropriately assume cognitive responsibility for a task, and the developmental consequences of that assumption for a learner. SCAN operates at each task completion process by “scanning” the human-AI task assignment by task identification, task processing, and task evaluation. In what follows, we introduce two core components in the SCAN framework: Task Paradigm and Metacognition.

## 2.1. Task Paradigm

There are four subzones in the SCAN's Task Paradigm (Figure 1). In the *Substitute* subzone, AI automates the task as a learner has no task-specific knowledge to complete with. The task does not draw on the learner's judgment, and its output cannot be independently verified by the learner at this stage. In contrast, AI collaborates with a more expert learner to extend one's own capacity beyond what unaided human cognition could achieve in the *Complement* subzone. The task requires both established learner expertise and AI capability, and neither alone is sufficient. In the *Aid* subzone, AI augments a learner who retains active cognitive ownership. The task requires learner judgment. AI provides support as “a digital scaffold” that a learner, with epistemic humility, critically evaluates and integrates. The interaction is designed to extend, rather than replace, the human's reasoning (Cabitza & Vicente, 2026; Tran et al., 2025). In the *Non-Negotiable* subzone, augmentation occurs through human experts as MKOs rather than AI. The task requires the kind of relational, ethical, or embodied judgment that AI cannot replicate, or the developmental cost of AI automation is unacceptable (Guth et al., 2024; Ruczynski et al., 2022; Wood et al., 1976).

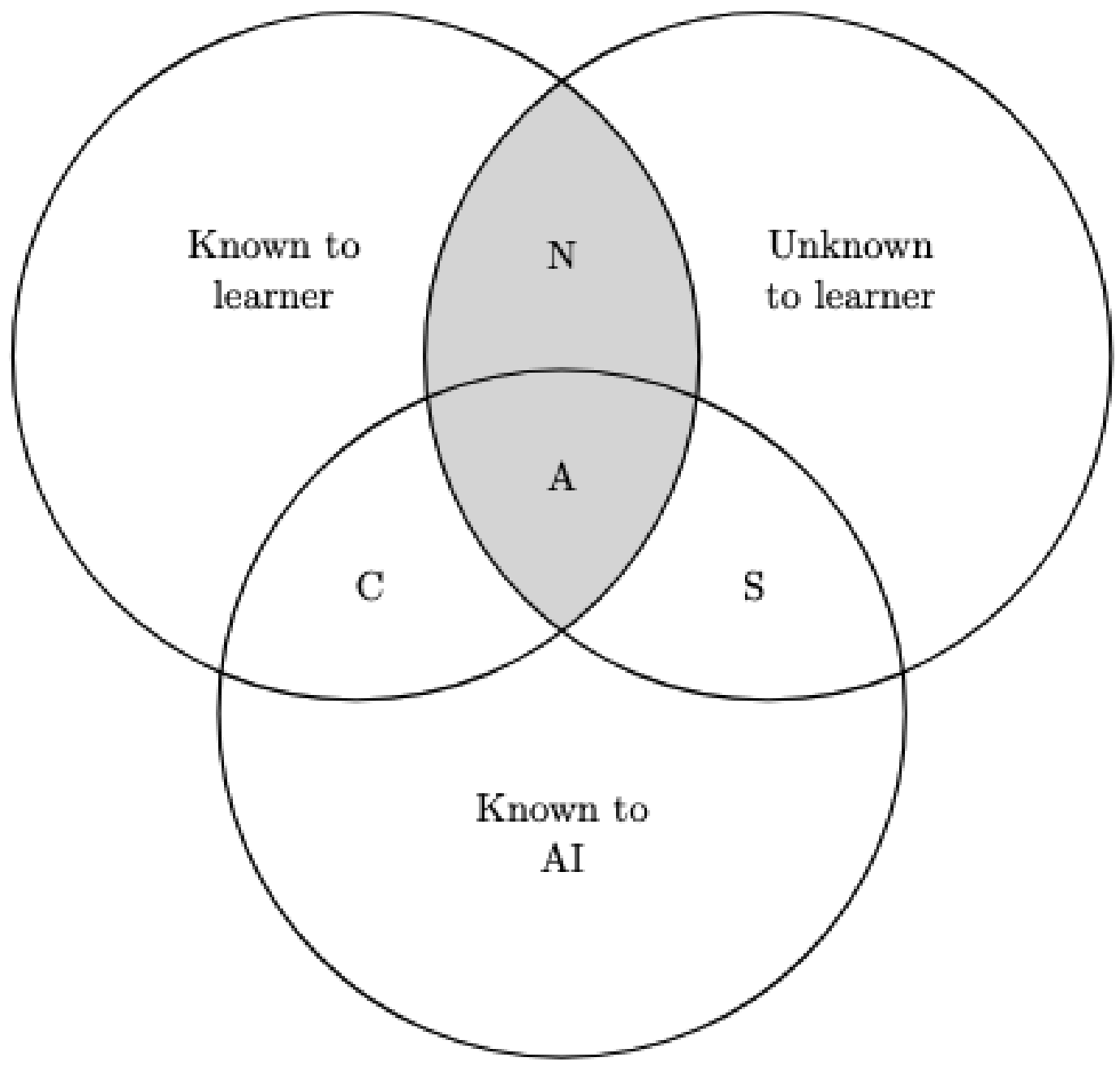

*Figure 1 Task Paradigm. SCAN extends Vygotsky's ZPD (shaded region) which is located in the overlapping area of two circles: "known to learner" and "unknown to learner," indicating zones of actual and potential development, by adding a new circle "known to AI". There are four overlapping regions marked as S (Substitute; between "unknown to learner" and "known to AI"), C (Complement; between "known to learner" and "known to AI"), A (Aid; the overlapping area of all three circles), and N (Non-negotiable; original ZPD). Retrieved from Tsim & Gutoreva (2026).*

### 2.2. Metacognition

Broadly defined as one's awareness and regulation of one's own cognitive processes (Cox, 2005; Flavell, 1979; Schraw & Moshman, 1995), metacognition is the second structural component of SCAN and its primary operational mechanism. Whereas SCAN's Task Paradigm delineates where a given task sits relative to a learner's current competence boundary, metacognition governs how the learner perceives, monitors, and adjusts that boundary in real time. Without adequate metacognition, task classification collapses: a medical student who cannot accurately evaluate their own knowledge state cannot reliably assign tasks to the correct SCAN subzones. In AI-mediated medical environments this failure is compounded: recent work shows that as reliance on AI increases, accuracy in detecting AI errors decreases—yet user confidence does not (Fernandes et al., 2026; Klingbeil et al., 2024). This performance–metacognition dissociation is particularly consequential in clinical training, where misplaced confidence can propagate downstream into patient care.

Effective use of generative AI requires not only domain competence but a deliberate, reflective metacognitive evaluation of AI use against one's own epistemic boundaries — a process structurally analogous to the mindful body scan, in which systematic, non-judgemental attention is directed to cognitive activity (Carmody & Baer, 2008). This capacity to assess the reliability and failure modes of AI-generated outputs constitutes what SCAN operationalises as metacognitive zone identification (Bergamaschi Ganapini et al., 2025; Tankelevitch et al., 2024). Absent this capacity, learners default either to over-reliance—automation bias—or to avoidance, both of which impair skill acquisition and clinical performance (Pham et al., 2025; Izquierdo-Condoy et al., 2025). Flavell's (1979) foundational account organized metacognition around three interacting components—metacognitive knowledge, metacognitive monitoring, and metacognitive control—each mapping onto a distinct functional role within SCAN.

#### 2.2.1. Metacognitive Knowledge

Metacognitive knowledge refers to what a learner knows or believes about themselves as an individual, about the tasks they face, and about the strategies available to address those tasks (Flavell, 1979; Schraw & Moshman, 1995). Critically, this knowledge is not static: in AI-mediated learning environments, it must continuously expand to include calibrated models of machine capability and failure modes, updating through iterative experience with AI-generated outputs, which underlies the essence of learning (Kasneci et al., 2023). Within SCAN, this constitutes the representational precondition for initial zone identification: the judgment that a pending task belongs in Substitute, Aid, or Complement. In clinical training, this judgment is anticipatory—it relies on domain specialization and prior experience to select cognitive policies before the task begins (Versteeg et al., 2021). An experienced clinician rapidly recognises which elements of a differential diagnosis she can generate independently (Complement) and which require AI-assisted retrieval across a larger evidence base (Aid); a first-year medical student often cannot make this distinction reliably, because the underlying schema does not yet exist.

Empirically, deficits in metacognitive knowledge are both pervasive and consequential among novice clinical learners. A landmark study by Cleary et al. (2019) examined 157 first-year medical students engaged in virtual patient simulations: across history-taking and physical examination subtasks, 95–98% of students overestimated their own performance, producing a systematic calibration bias that was more pronounced for the cognitively demanding subtask. Garbayo et al. (2023) replicated this pattern in high-fidelity patient simulations (n = 80 teams): faculty-rated diagnostic accuracy ranged from 23–74%, while student self-reported confidence ranged from 71–86%. Crucially, team discussion narrowed—but did not close—this gap. These findings identify the misclassification problem SCAN is designed to address: when metacognitive knowledge is inaccurate, learners place Aid-level tasks in the Substitute zone, delegating clinical judgment they do not yet have the competence to supervise. In AI-mediated environments, this tendency is exacerbated by the surface fluency of large language model outputs, which novices frequently conflate with epistemic reliability (Colville & Ostern, 2026; Safarov et al., 2026).

### 2.2.2. Metacognitive Monitoring

Metacognitive monitoring refers to the ongoing, real-time process by which learners track the quality of their understanding and performance during task execution (Nelson & Narens, 1990; Flavell, 1979). Within SCAN, monitoring constitutes real-time evaluation: the continuous assessment of whether the current task remains appropriately situated in its designated zone or requires immediate reassignment. A medical student working on a differential diagnosis who notices her reasoning is circular—drawing inferences from the same two data points without expanding the hypothesis space—should, if her monitoring is functioning, recognise this as an Aid signal and invoke AI-assisted literature retrieval rather than proceeding to Substitute-level delegation.

Studies in medical education consistently find that metacognitive monitoring is underdeveloped relative to domain knowledge acquisition, and that stronger monitoring accuracy predicts superior diagnostic reasoning over time (Wang et al., 2023; Yan et al., 2026). Critically, the verification bottleneck documented in AI-assisted environments—where accuracy in detecting AI errors decreases even as confidence holds steady—is precisely a failure of monitoring (Fernandes et al., 2026; Klingbeil et al., 2024): the learner's sensing layer is not registering the growing gap between machine output reliability and her own supervisory capacity. Interventions that make monitoring explicit—such as the metacognitive confidence calibration tool trialled by Garbayo et al. (2023) in high-fidelity simulations—have demonstrated improved quality of deliberate reasoning in novices and represent a direct clinical instantiation of what SCAN's real-time evaluation mechanism requires (Wang & Huang, 2026) .

### 2.2.3. Metacognitive Control

Metacognitive control—also termed metacognitive regulation—refers to the learner's capacity to adjust cognitive strategies and resource allocation in response to monitoring outputs (Flavell, 1979; Nelson & Narens, 1990; Schraw & Dennison, 1994). Within SCAN, control operates across two temporal scales. Prospectively, it governs zone selection prior to task engagement: whether to delegate, scaffold, or proceed independently. Retrospectively, it governs reflection: post-task evaluation of whether the assigned zone was appropriate, whether AI-generated outputs contained detectable errors, and what adjustments are warranted for similar tasks in the future (Panadero, 2017; Zimmerman, 2002). In clinical education, this reflective function is central: a first-year resident who delegates a discharge summary to a generative AI system, for example, reviews the output without scrutiny, and submits it has enacted Substitute-level control over a task she classified as Complement—a zone-behaviour misalignment that SCAN's control mechanism is designed to prevent.

Taken together, these knowledge, monitoring and control components constitute a unified regulatory architecture that SCAN operationalises across the clinical task lifecycle. Metacognitive knowledge frames the initial zone identification judgment; monitoring executes it in real time; and control adjusts it retrospectively, enabling the progressive movement from other-regulation toward the self-regulation that clinical expertise requires (Wang et al., 2023; Bergamaschi Ganapini et al., 2025). Importantly, this architecture is not inherently threatened by AI: well-designed human–AI interaction can itself serve as a metacognitive scaffold, prompting learners to make their reasoning explicit and engage in deliberate self-monitoring that unassisted study rarely elicits (Burgos-Martínez et al., 2026; Gao et al., 2026; Tran et al., 2025). SCAN operationalizes this scaffold by making zone assignment—and its metacognitive justification—an explicit, auditable step in the clinical learning process.

## 3. Clinical Reasoning Development

### 3.1. Two Developmental Trajectories in the Age of AI

Understanding how AI disrupts clinical reasoning development requires first establishing what development looks like under SCAN. We believe it accounts for two distinct developmental trajectories simultaneously. The first one is the *Internalization Trajectory*, which preserves Vygotsky's original account of human development through the Non-negotiable subzone with human MKO. In clinical training, the apprenticeship model has historically instantiated this Vygotskian logic (Bleakley, 2006). Motivated by the growth-oriented intentions for clinical reasoning development, the junior trainee observes the expert, then participates with supervision, then performs independently with decreasing oversight (Wei et al., 2026). Each stage requires that the trainee actively engage with the cognitive demands of the task, by developing the underlying reasoning capacity that would allow them to produce correct outputs across novel situations.

Further, SCAN proposes the *Symbiosis Trajectory* where a learner navigates from Substitute through Aid to Complement, progressively developing task-specific knowledge in collaboration with GenAI (Tsim & Gutoreva, 2026). In the Substitute subzone, the learner engages in *other-regulation*: task-specific knowledge is absent, and the learner cannot verify or direct AI output. A first-week student generates an initial differential diagnosis with AI for an unfamiliar presentation, as the cognitive work exceeds current capacity for critical engagement on diagnostic accuracy (Teng et al., 2026). In the Aid subzone, partial task-specific knowledge enables *shared regulation*: a third-year student clerking a chest pain case can augment one's own reasoning with AI-generated suggestions, and question outputs that conflict with observed clinical findings (Tang et al., 2026). In the Complement subzone, sufficient task-specific knowledge enables *self regulation*: a senior resident directs LLMs to draft sections of a complex discharge plan, critically evaluates the output against their own clinical judgment, and takes full epistemic responsibility for the final document (Raghu Subramanian & Rosner, 2025). This form of self-regulation differs qualitatively from the one Vygotsky described, in which the external governance of the AI interaction has been fully internalized as the learner's own cognitive practice.

In short, these two trajectories constitute the full developmental architecture of clinical training with the recent, rapid development and incorporation of GenAI. On the one hand, the Internalization Trajectory produces the tacit, situated expertise that defines clinical mastery. On the other hand, the Symbiosis Trajectory produces a qualitatively different but equally necessary competence: the metacognitive sophistication to direct, evaluate, and take epistemic responsibility for AI as "an extended self" (Clark & Chamber, 1998; Gutoreva et al., 2026). Neither trajectory is optional, and neither substitutes for the other. We conjecture that complete clinical formation requires both endpoints (Kassab et al., 2025; Stamos & Dharod, 2026). Both share the same normative demand: epistemic agency (Samuel, 2025) and epistemic responsibility (Schwab, 2008) must increase progressively as the learner develops.

### 3.2. Misclassification: the primary disruption to clinical reasoning development

Within the Symbiosis Trajectory, clinical reasoning development can be disrupted through misclassification, which is defined as selecting a subzone-inappropriate AI interaction mode for a given task at a given stage of development. We identify there are two distinct paths of misclassification located at the Aid subzone, with each producing a different failure mode due to different psychological mechanisms.

#### 3.2.1. Bypassing productive cognitive engagement

Suppose a learner treats a task that sits within the Aid subzone (where augmentation occurs: productive struggle is both possible and necessary for development) as a Substitute one (where automation occurs). This action could be attributed to two potential causes. First, it is due to *effort minimization* (Misiejuk et al., 2026): a learner defaults to automation as it is the path of the least resistance. This is a clear sign when a response of mental shortcut ("System 1") bypasses the deliberate thinking ("System 2") that zone identification requires (Kahneman, 2003, Kahneman, 2011). Secondly, it is due to *authority heuristic*: the presence of an AI system carries an epistemic authority signal, leading to acceptance that the AI output is more reliable than one's own judgment (Cialdini & Goldstein, 2004; Mosier et al., 1998; Parasuraman & Manzey, 2010). The learner is susceptible to this heuristic due to partial knowledge to engage the task in the Aid subzone. With a failure of real-time metacognitive evaluation (Ackerman & Thompson, 2017) due to these two psychological mechanisms, a learner eliminates the cognitive encounter through which relevant clinical skills would form otherwise.

### 3.2.2. Cultivating an illusion of competence

In the second path, a learner treats an Aid task as belonging to the Complement subzone by engaging with AI as a "proficient" delegator, without possessing the task-specific knowledge that Complement task requires. This action could be attributed to *overconfidence* driven by the Dunning-Kruger effect (Kruger & Dunning, 1999): a learner's insufficient task-specific knowledge produces an inflated self-assessment due to lack of metacognitive awareness to recognize the limit of their knowledge (Al Bitar & Chen, 2026). Unlike the first path where the learner underutilizes their partial knowledge, the learner in this path overestimates it as a complete one (miscalibrated engagement). AI errors, which sounds plausible, pass through verification that feels rigorous but is not, as the sufficient level of knowledge required to detect those errors is exactly what the learner does not yet possess (Teng et al., 2026).

In terms of clinical reasoning development, misclassification disrupts learner's progression in two distinct ways, which existing framings of AI overreliance or misuse fails to capture. In the first path of misclassification, treating an Aid task as a Substitute task eliminates the cognitive struggle that could have been formed in the Aid task engagement. In the second path, treating an Aid task as a Complement task provides "pusedo-competence" (Ding & Weng, 2026), or an illusion that the learner possesses sufficient task-specific knowledge to direct and evaluate AI output with confidence. Succinctly put, the first path prevents the learner from progressing, and the second path produces a false impression that the learner has progressed.

Apart from that, the distinction of Misclassification from Misuse is significant in terms of potential causes and solutions in terms of curriculum design. On the one hand, misuse implies a norm violation: the learner knew or should have known what the correct behaviour was, and deviated from it. As a result, the natural response to misuse is *enforcement*: sanctions, restrictions, monitoring (Knopf et al., 2026). A curriculum organized around misuse prevention will invest in policy, surveillance, and enforcement. On the other hand, misclassification, with its two pathways yielding distinguishable and falsifiable predictions, implies a competency deficit: the learner lacked the metacognitive monitoring skill to evaluate their own AI interaction appropriately (Flavell, 1979). The natural response to misclassification is *instruction*: explicit teaching of subzone classification as a cognitive skill, with scaffolded practices in making classification decisions, and formative feedback on classification accuracy, which in turn, fostering learner's metacognitive ability or "epistemic cognition" (Barzilai & Zohar, 2016; Chang et al., 2021; Eastwood et al., 2017). A curriculum organized around misclassification prevention, which targets to mitigate related psychological mechanisms, will invest in metacognitive instruction, task design, and supervised practice. It is thus both more educationally coherent, and more likely to produce lasting effects, as it addresses the underlying cognitive failure rather than merely constraining its expression.

### 3.3. Trajectories of Skill Acquisition and Failure

In this section, we examine two distinct consequences of task-subzone classification in AI-assisted clinical training. The first one is upskilling, which is the result of correct engagement driving tasks identification from Substitute to Aid, then from Aid to Complement as knowledge accumulates over time. The second one is "The Triad of Skill Failure", which are the harms that arise when the developmental process is disrupted. Misclassification drives never-skilling and mis-skilling partly (depending on which zone boundary is crossed), and deskilling arises independently through regression. Importantly, some elements in the triad might not constitute harm. Ke et al. (2026) note that surgeons trained after the widespread adoption of laparoscopy were never taught to operate without it (never-skilling); they do not need to. The questions, we argue, are whether the underlying task subzone classification was correct at the start, and whether it is followed with appropriate subzone engagement.

#### 3.3.1. Upskilling and the metacognitive learning component as the counterpart

In clinical reasoning development, we argue that routine clinical tasks, properly scaffolded by metacognitive self-evaluation, are *subzone-migration opportunities*. This reframes these tasks as "developmentally generative" rather than uniformly offloadable or uniformly indispensable: its value is *subzone-dependent* and *metacognitively governed* with the following three metacognitive components. First, Metacognitive Monitoring determines current subzone position before the AI interaction begins. Its accuracy is the primary safeguard against misclassification, and the primary target for faculty calibration in early training (Yan et al., 2026). Secondly, Metacognitive Control operates retrospectively after task completion (Daniel et al., 2026). For instance, a learner asked: “was the interaction mode appropriate? What did the AI output contain that could and could not be verified?” This retrospective interrogation itself is a form of clinical reasoning practice. This can occur even with Substitute tasks, as the learner must evaluate the quality of what was produced. Finally, learning is the component that drives subzone migration over time. Tasks should migrate from Substitute toward Aid then Complement as task-specific knowledge accumulates. Such migration is the observable sign that internalization is occurring.

Upskilling is an upward, proactive shift in the SCAN's Task Paradigm with learner's metacognition. It requires a learner to re-evaluate one's own knowledge state in the moment, and update one's own interaction mode accordingly (*learner-dependent*; Gao et al., 2026). As a result, this offers an unambiguous, systematic pathway from a novice to an intermediate, from an intermediate to become an expert (where mastery is thus attained).

### 3.3.2. The Triad of Skill Failure in Clinical Training

Misclassification is not a single type of developmental harm in AI-assisted clinical training. An emerging literature investigated three other types of AI-related skills failure, or known as "The Triad of Skill Failure" (Abdulnour et al., 2025; Berzin & Topol, 2025; Ke et al., 2026; Keren, Desai & West, 2026). In what follows, we identify three corresponding, directionally distinct pathways in the SCAN's Task Paradigm — each with different mechanisms, different recoverability, and different intervention requirements (Figure 2, Table 2).

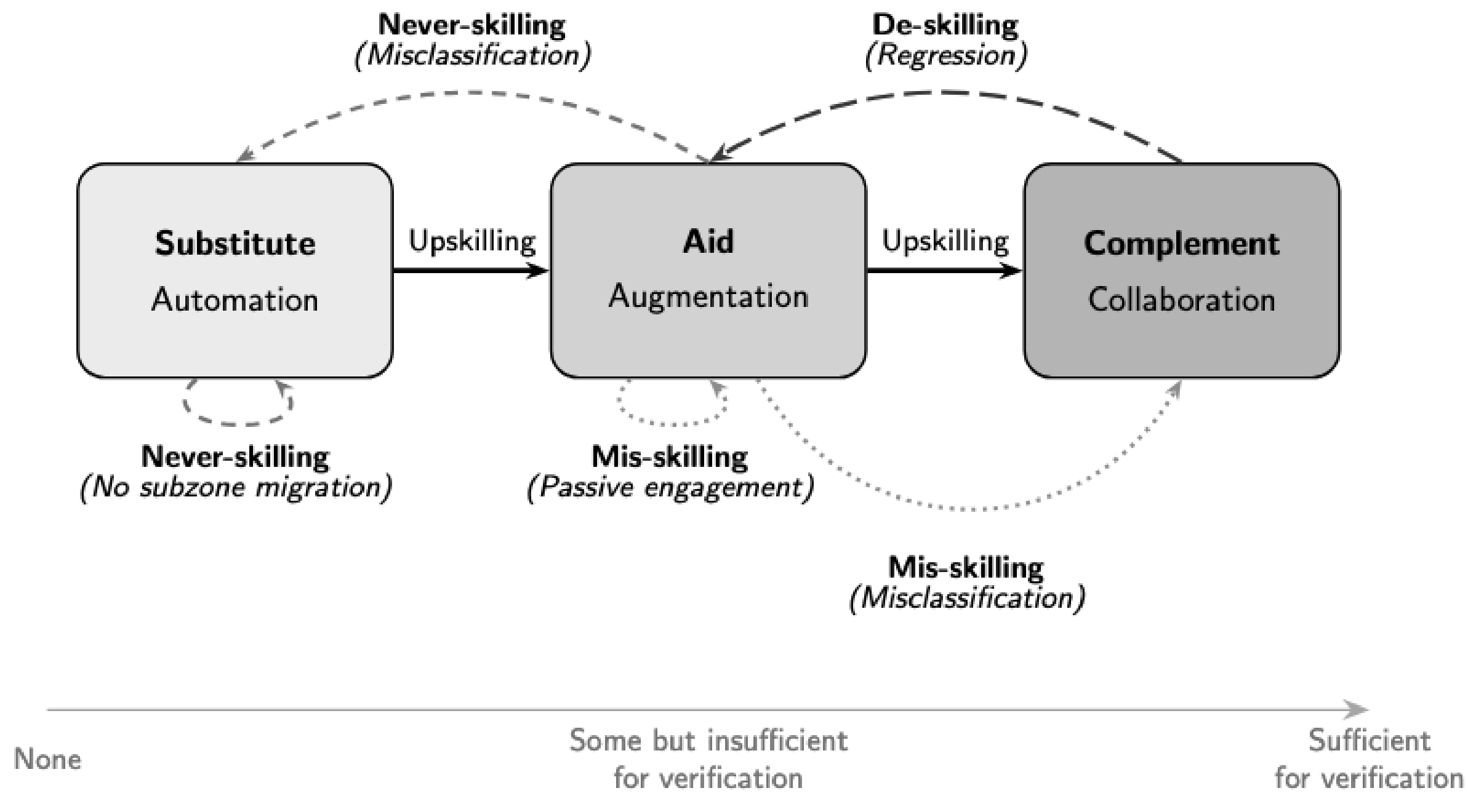

*Figure 2 Mapping Upskilling and the Triad of Skill Failure (De-skilling, Never-skilling, and Mis-skilling) in the SCAN's Task Paradigm (Substitute, Aid, and Complement, corresponding to human-AI decision making modes (Automation, Augmentation and Collaboration)). Upskilling is the migration of tasks from Substitute toward Aid and Complement as a learner's task-specific knowledge accumulates through correct, subzone-appropriate engagement. Deskilling is the regression of a Complement task to Aid. Never-skilling and mis-skilling each arise through two pathways: (1) a misclassification pathway where an Aid task is wrongly treated as Substitute or Complement (dashed arcs), and (2) a non-misclassification pathway where the subzone is correctly identified but the process still fails — no subzone migration within Substitute, or passive engagement within Aid (dotted loops).*

#### 3.3.2.1. Deskilling

*De-skilling* refers to the atrophy of a skill that the learner previously possessed, through sustained AI-mediated disuse (Monteith et al., 2026; Natali et al., 2025). This atrophy is driven by the repeated exercise of automation bias — each individual act of delegation may be reasonable, even efficient, but its accumulation, over time, erodes the very skill being delegated (Heudel et al., 2026). According to SCAN, it occurs when a task that was engaged as at the Complement subzone is now identified regressively to the Aid subzone only (the learner retains residual task-specific knowledge, but gradually loses the fluency and depth that self-regulation has achieved). The regulatory regression is from self-regulation back toward shared regulation. De-skilling is a *retrospective* harm for development: it degrades the skill that already exists due to lack of practices. Among the triad, its intervention is the most straightforward in principle: reinstating the practice that was displaced. Also, de-skilling is the most visible failure mode among all, as the deterioration in performance assessments that previously showed competence is visible, and thus detectable (Levartovsky & Kopylov, 2025).

##### 3.3.2.2. Never-skilling

*Never-skilling*, or known as "upskilling inhibition", is defined as the failure to develop a skill in the first place, as AI works as a replacement at the formative stage before the skill has been established (Ke et al., 2026; Natali et al., 2025). Direct evidence for never-skilling in clinical trainees remains absent. Ke et al. (2026) present it as a conceptual risk model with three interrelated proposed mechanisms (competency acquisition failure, calibration paradox, and metacognitive erosion) grounded in adjacent, non-clinical signals rather than an established clinical phenomenon. In what follows, we adopt the same caution here. SCAN shows that never-skilling has two potential pathways. First, the misclassification pathway is when the learner misclassifies an Aid task as a Substitute one, due to a failure of real-time metacognitive monitoring that eliminates the cognitive struggle before any competency can form. We recognize it maps onto Ke et al's (2026) competency acquisition failure. Next, the non-misclassification pathway is when learners identify the substitute task correctly, but the metacognitive learning component fails to trigger subzone migration over time, so the learner remains in the Substitute subzone indefinitely. The other two mechanisms of Ke et al's (2026) describe downstream, cumulative consequences of substitution rather than a distinct origin, and neither accounts for the non-misclassification pathway.

We recognize SCAN's pathway-level precision comes at a cost relative to Ke et al.'s (2026) population-level, phased framework. Their account offers ready-made, cohort-wide curricular policy and better captures the cumulative, identity-level drift from sustained AI exposure. SCAN's task-level account, in contrast, captures within-cohort heterogeneity: a learner may be past AI-free training on one task while still requiring it on another. SCAN's approach, which tracks classification accuracy at the task-level for a learner, implies different interventions for the two pathways. The misclassification pathway calls for improved calibration of real-time self-assessment accuracy (learner's use of the metacognitive monitoring (real-time evaluation) and control (reflection) components). The non-misclassification pathway, which could lead to no developmental harm, calls for structured subzone-migration checkpoints (and thus trigger learners' use of the metacognitive learning component, and prompt re-evaluation of the knowledge state). Given the absence of direct evidence for never-skilling, what we offer here are mechanistic, falsifiable hypotheses for future measurements and interventions rather than settled claims — the terms on which SCAN's task-level account should, with Ke et al.'s population-level account, be investigated.

In short, never-skilling can be a *prospective* harm for development: it prevents the skills that should be formed. Compared to de-skilling, we conjecture it is less visible as the absence of a skill that was never present in the first place does not register as deterioration in the assessments that assume prior competence.

#### 3.3.2.3. Mis-skilling

*Mis-skilling* is denoted as the internalization of incorrect, biased, or contextually inappropriate clinical schemas derived from AI outputs (Teng et al., 2026). According to SCAN, it is located at the Aid-subzone, where a learner has some task-specific knowledge but insufficient for verifying AI outputs. Inability to verify AI output due to insufficient knowledge, makes critical engagement with AI output unreliable. Plausible-sounding errors pass undetected, and faulty clinical reasoning forms through the interaction rather than despite it (Obermeyer et al., 2019). Mis-skilling has two potential pathways: (1) learners passively engage with an Aid task (a failure of *epistemic activity*), and (2) when misclassification happens: the learner misclassifies an Aid task as a Complement one, due to one's own overconfidence on one's own level of task specific knowledge (Dunning-Kruger Effect). Mis-skilling is a *constructive* harm for development: it actively builds wrong clinical structures that may need to be unlearned. This connects to the literature on unlearning in cognitive science: unlearning is harder than learning from scratch, as wrong structures must be inhibited before correct ones can consolidate (Bjork & Bjork, 1992; Kendeou & O'Brien, 2014). Mis-skilling is, we conjecture, a dangerous failure mode in the triad, as its directionality is constructive rather than degenerative or preventive. It does not announce itself as an absence; it presents itself as learning.

*Table 2. Three types of AI-related skills failure modes.*

| | Deskilling | Neverskilling | Mis-skilling |
|---|---|---|---|
| **Location** | Complement → Aid | (1) Aid → Substitute (Misclassification); (2) Substitute | (1) Aid → Complement (Misclassification); (2) Aid |
| **Type of harm** | Retrospective (Degradation) | Prospective (Prevention) | Constructive (Construction) |
| **Current Empirical Support** | Direct clinical evidence (Budzyń et al., 2025); no medical education specific study yet exists | None; conceptual risk model only (Ke et al., 2026) | Direct RCT evidence (Teng et al., 2026) |

An important, yet underappreciated, form of mis-skilling involves not subzone selection but engagement quality within a correctly selected subzone (a failure of *epistemic activity*). Suppose, with a given development stage, a trainee correctly identifies a clinical documentation task to the Aid subzone, but engages passively with the AI output (e.g., reading without evaluating, accepting without comparing against their own reasoning). AI-generated content is processed and potentially internalized, but the incomplete knowledge base means plausible-sounding errors pass undetected, and the cognitive interaction that makes Aid-subzone engagement developmentally beneficial fails to instantiate. As a result, wrong clinical structures form through the interaction rather than despite it. In short, passive engagement at the Aid subzone poses a fundamental practical challenge for clinical supervision. Within a correctly identified Aid task, a learner passively engaging with AI is indistinguishable from the one exercising actively. The solution requires a "subzone re-identification" from Aid to Non-Negotiable, whereby human experts serve as epistemic auditors examining not which mode the learner selected, but how the mode engagement within that looked like. This subzone re-identification aligns with Vygotsky's ZPD framework: where AI cannot ensure active cognitive participation, humans as MKOs can.

## 4. Applications

### 4.1. Curriculum design

A SCAN-informed curriculum treats subzone classification as a teachable and assessable cognitive skill. This requires three design commitments. First, clinical curricula must include explicit instructions on the four SCAN subzones in order to orient the learner to this theoretical framework and allow them to appropriately apply it to a task. While the concepts of deskilling, mis-skilling and neverskilling have recently been introduced in the medical education literature (Bowen & Abdulnour, 2025; Keren et al., 2026), they have not yet been widely incorporated into curricular design in clinical courses and training programs.

Second, the learner must have an understanding that assignment of a clinical task to a particular subzone varies based on the level of a learner as they progress from novice to expert. Consider, for example, the generation of a differential diagnosis for a middle aged patient in the emergency department presenting with new onset chest pain. A second year-medical student on the first day of their emergency medicine rotation assigned to care for this patient likely has only minimal prior knowledge and experience with this type of case. Because of this, allowing AI to perform the task of developing the differential diagnosis should be a non-negotiable task (guided by human experts) and thus AI-free. In contrast, a senior emergency medicine resident assigned to this patient has likely seen hundreds of similar cases of chest pain previously, and has had formal instruction in their core residency curriculum on the topic. Therefore, this resident may use AI as a scaffold to develop the differential diagnosis without hindering the development of their clinical reasoning ability.

Third, it is imperative that faculty development initiatives are in place in order for clinical educators to understand their role as *calibrators* of learners' subzone-classification accuracy, particularly in early training when self-assessment is weakest and never-skilling risk is highest (da Fonseca Rezende et al., 2025; Ke et al., 2026). Faculty must be supported and empowered to calibrate their learners' subzone classification and make subzone boundaries visible and discussable, which is in line with current guidelines and recommendations and for the appropriate use of AI in the clinical learning environment (Association of American Medical Colleges , 2025; Gin et al., 2025).

### 4.2. Clinical supervision

The SCAN framework offers clinical supervisors a structured vocabulary for a conversation that has previously lacked one. When a trainee delegates a documentation task to AI, the supervisor currently has limited conceptual resources for evaluating that decision: it is either permitted or not permitted, depending on institutional policy. SCAN provides a richer evaluative frame: the question is not whether the delegation occurred but whether the subzone classification was appropriate, and more importantly, the engagement quality was active. SCAN-informed supervision requires two explicit faculty actions. The first is zone verification: the supervisor asks the trainee to articulate and justify their zone assignment — why this task was treated as Non-negotiable (instead of Substitute and Aid) or Complement given their current fund of expertise with the task. This surfaces miscalibration before it consolidates into habit. The second is output interrogation: the supervisor asks the trainee not merely whether the AI output is correct, but whether they can reconstruct the reasoning behind it, identify its failure modes, and integrate it with the broader clinical picture. Together, these two moves shift supervision from output evaluation to process evaluation — the core diagnostic shift SCAN proposes. This mirrors established supervision models in which questioning the learner's reasoning process, rather than auditing outputs, is the primary teaching act (ten Cate et al., 2015; Kilminster & Jolly, 2000).

A substantial body of empirical evidence demonstrates that medical trainees systematically overestimate their own competence, particularly at early stages of training when task-specific knowledge is most limited. Studies across multiple specialties and training levels have documented significant discrepancies between self-assessed and externally evaluated performance, with overconfidence most pronounced precisely among the least experienced learners (da Fonseca Rezende et al., 2025; Fleming et al., 2021; Knof et al., 2024; Kuhn et al., 2022). This pattern is consistent with the Dunning-Kruger effect, wherein limited domain knowledge impairs the metacognitive capacity needed to accurately evaluate one's own performance  producing a double deficit in which the learner both lacks competence and lacks the awareness to recognise that lack (Kruger & Dunning, 1999; Rahmani, 2020). Although the specific capacity of a learner to make an accurate SCAN subzone assignment has not yet been empirically tested, this literature warrants serious concern: if trainees routinely overestimate their clinical competence in general, they may be equally prone to misclassifying their position within the SCAN framework — assigning themselves to Aid or Complement when their actual knowledge state places them firmly in Substitute. The consequences of such misclassification are not trivial. A trainee who incorrectly self-assigns to Complement and delegates an Aid-level task to AI without adequate verification lacks the task-specific knowledge required to detect errors in AI output, a failure mode we term mis-skilling (see Figure 2). This theoretical concern is reinforced by evidence that miscalibration is resistant to correction: even targeted feedback on diagnostic accuracy does not reliably improve self-assessment on more complex clinical cases (Kuhn et al., 2022), suggesting that awareness alone is insufficient. Faculty supervision is therefore not merely a pedagogical preference but a structural necessity within the SCAN model. The attending physician functions as an external calibrator — the more knowledgeable other in Vygotskian terms (Vygotsky, 1978) — whose role is to verify zone assignments that the trainee cannot yet verify independently, and to provide the corrective feedback that self-directed metacognitive monitoring cannot reliably generate at early stages of expertise development (Davis et al., 2006; Eva & Regehr, 2008). For instance, when a trainee delegates a chest-pain differential diagnosis to AI, zone verification asks whether their knowledge state supports Aid-level engagement; when they delegate medication reconciliation, output interrogation probes whether they can detect interaction errors the AI may have missed (Schiff et al., 2009). It's important to note that not all offloading is developmentally harmful: delegating formatting of a discharge summary a trainee has already mastered carries no cost (appropriate Substitute), whereas delegating the clinical reasoning within it does (misclassified Aid). The supervisory task is distinguishing the two (Ericsson, 2004).

Some delegation may actually free up the dyad for higher-level pedagogical engagement, for example, using AI to offload activities known as "cognitive SCUT" (da Fonseca Rezende et al., 2025; Fleming et al., 2021; Kelly et al., 2025; Fischer Lees et al., 2023; Oxentenko et al., 2010). The high-volume, routine clinical tasks such as documentation, order entry, reviewing results, or preparing discharge instructions, for example, carry significant, yet under-appreciated developmental value. AI, therefore, makes precisely these tasks the most easily "offloadable", thereby posing risks to clinical reasoning development, which current frameworks have not yet differentiated by learner knowledge state or task-specific developmental value (Knof et al., 2024; Kuhn et al., 2022).

### 4.3. Assessment

Existing approaches to clinical reasoning assessment in AI-assisted training share a limitation, that is, they evaluate primarily through outputs rather than processes. Outcome assessment, unlike process assessment, cannot distinguish behaviorally a student's output that is either from an appropriate subzone classification or misclassification. For instance, in a recent investigation of attending and resident physicians trained in internal medicine, family medicine, or emergency medicine, those who used a LLM for diagnostic reasoning (a classification task with single correct answer), did not show significant improvement over those using conventional resources alone (Goh et al., 2024). For instance, a closed diagnostic task such as identifying new-onset type 1 diabetes from a classic presentation offers little for an LLM interaction to disrupt or enhance, since the single correct answer is reachable by pattern recognition alone, whether the physician's own or the model's. Apart from that, when it comes to management reasoning (contextual, patient-specific decisions with no single correct answer), physicians who use LLM significantly outperformed and spent more time per case than those without it (Goh et al., 2025). A management task such as anticoagulation planning in a patient with competing bleeding and stroke risk has no single correct answer, requiring the physician to weigh AI output against patient-specific context the model cannot access, which plausibly explains both the performance gain and the added time. These divergent findings, through the SCAN framework, reflect a subzone classification difference. Diagnostic tasks appear to have triggered passive engagement and offloading (mis-skilling), while management tasks prompted Aid-subzone engagement. Process assessment, which is structured around the gating condition that classification accuracy must be established before the interaction quality can be evaluated meaningfully, makes the mechanism visible. In what follows we discuss how process-focused assessment is operationalized in practice.

#### 4.3.1. Two-layer assessment architecture

First, the assessment within SCAN operates in a visible sequence. The first layer evaluates subzone classification accuracy: did the learner identify the task correctly based on their current knowledge? Once the task is classified correctly, the second layer evaluates AI interaction quality within the subzone. Otherwise (where misclassification happens), a misclassified task produces either no engagement to evaluate (never-skilling), or engagement that appears correct to surface inspection (mis-skilling). The visibility in the classification decision requires externalizing the internal metacognitive event. For instance, a written log spanning three moments (before, during and after the interaction) provides a traceable process record that is independent of clinical outcome, which maps onto the three metacognitive components that SCAN requires.

#### 4.3.2. Subzone-sensitive distribution of responsibility

Secondly, the assessment responsibility between an instructor and a learner varies across subzones. This can be done via, and testified with, subzone-classification accuracy that is tracked longitudinally through the written log. Instructor assessment is primary for Substitute tasks: learners lack the task-specific knowledge to verify their own classification decisions. One of the misclassification pathways can occur without learner recognition for Aid tasks. This, therefore, requires the instructor to function as epistemic auditors of the interaction process rather than outcome. For Complement tasks, sufficient knowledge enables meaningful learner self-assessment, and thus instructor oversight shifts significantly from corrective to confirmatory. Put it succinctly, this distribution of responsibility implies a sequencing principle: instructor involvement should precede learner interaction with AI. It gradually withdraws as learner's task-specific knowledge accumulates through AI interactions over time, consistent with both Internalization (human-human learning; Vygotsky, 1978) and Symbiosis Trajectories (human-AI learning; Tsim & Gutoreva, 2026).

## 5. Research Agenda

With the theoretical foundation and conceptual reframing proposed in this paper, we suggest four questions about the prediction of clinical reasoning quality, the mechanism-specific remediation of misclassification's two pathways, the detection via passive engagement in the Aid task of, and exploring behavioral interventions that could lead to remediation for, mis-skilling.

### 5.1. Does metacognitive accuracy in zone classification predict the quality of clinical reasoning development?

The first question is a foundational one. If misclassification is a metacognitive failure, then metacognitive accuracy in zone classification should predict the quality of clinical reasoning development in AI-integrated training environments. This hypothesis can be tested through a prospective cohort study tracking medical students across the first three years of clinical training. Zone classification accuracy would be measured through a validated task-classification instrument at multiple time points; clinical reasoning development would be assessed through established measures including the Clinical Reasoning Examination and structured OSCE performance (Ilgen et al., 2012; Tekin et al., 2025). The prediction is that students with higher zone classification accuracy at baseline will show stronger clinical reasoning development over time. Also, changes in zone classification accuracy will precede, rather than follow, changes in clinical reasoning performance. Regarding novice learner's long-term development, the study would test whether the two never-skilling pathways ("no subzone migration from Substitute" and Misclassification) are distinguishable in their predictive signatures.

### 5.2. Does mechanism-specific intervention differentially remediates the two pathways of misclassification?

We introduce two distinct pathways of misclassification that disrupt their clinical reasoning development with related psychological mechanisms. If they are different mechanisms rather than variants of a single failure, they should respond differently to mechanism-matched intervention. This hypothesis can be tested through a randomized 2x2 trial using the task-classification instrument from the first agenda to identify each misclassified learner's pathway (leading to either never-skilling or mis-skilling), and random assignments of two targeted trainings. The prediction is that each training should reduce misclassification only among learners on its own targeted pathway, with no effect on the other. This "double dissociation" would rule out the possibility that either pathway is a milder variant of the other. A null result (i.e., uniform improvement regardless of pathway-training match) would falsify this paper's claim that the two pathways are mechanistically distinct.

### 5.3. Can passive engagement within correctly classified Aid tasks be detected before mis-skilling consolidates?

This question addresses the second pathway to mis-skilling, that is, passive engagement in the Aid task. This can be examined with a longitudinal observational study tracking learner engagement with Aid-zone AI interactions using process measures including think-aloud protocols (Tang et al., 2026), 'verify and trust' paradigm (Abdulnour et al., 2025), pre/post-consultation reasoning comparison tasks, and supervisory engagement ratings. This examination investigates whether passive engagement produces detectable signatures in reasoning performance before mis-skilling is established in assessment outcomes. Early detection is critical: mis-skilled reasoning that has consolidated connects to a well-established finding in the cognitive science of *unlearning* — inhibiting a wrong structure before correct ones can form is harder than learning from scratch (Bjork & Bjork, 1992; Kendeou & O'Brien, 2014), and the window for prevention may be narrow. This study can also benefit supervisors by generating the process indicators needed to train them in passive engagement detection.

### 5.4. Is mis-skilled clinical reasoning remediable, and through what interventions?

The third question is the most practically urgent. If mis-skilling produces reasoning schemas that are actively counterproductive and difficult to detect, we argue that the question of remediability is clinically significant: trainees who have been mis-skilled are underprepared and potentially unsafe. A mixed-methods intervention study (identifying trainees whose reasoning patterns show evidence of mis-skilling through the process measures developed in 5.3, then delivering and evaluating targeted remediation) would establish both whether mis-skilled reasoning is remediable or not, and if yes, then what behavioral intervention features are necessary (Bjork & Bjork, 1992; Kendeou & O'Brien, 2014).

The theoretical foundations for such an intervention already exist, though they have not yet been applied to AI-mediated clinical training. Cognitive remediation (CR) — broadly defined as behavioral training interventions designed to ameliorate deficits in cognitive functioning — has an established empirical record across neuropsychiatric populations, with meta-analyses demonstrating moderate to large effect sizes on both cognitive performance and functional outcomes in schizophrenia, mood disorders, and ADHD (Vita et al., 2021; Trapp et al., 2022). Critically, the literature identifies metacognitive strategy instruction — rather than drill-based task repetition alone — as the active ingredient that produces transfer to real-world functioning: CR interventions that explicitly teach monitoring and self-correction strategies consistently outperform those that rely on passive practice (Wykes et al., 2011; Trapp et al., 2022). This distinction maps directly onto the SCAN framework's remediation problem. Mis-skilling does not produce a knowledge deficit in the conventional sense; it produces a flawed reasoning schema that has been actively consolidated through repeated AI-mediated interaction.

Remediating it therefore requires not re-teaching domain content but targeting the metacognitive layer: rebuilding the monitoring and control processes that should have governed zone identification in the first place. Schema-level correction of this kind is structurally analogous to what CR achieves in neuropsychiatric contexts — inhibiting overlearned maladaptive patterns before correct ones can consolidate (Bjork, 1994; Fisher et al., 2011) — and the pedagogical principles that make CR effective (explicit strategy instruction, scaffolded practice with corrective feedback, transfer to ecologically valid tasks) translate directly to a SCAN-informed remediation curriculum. In medical education specifically, remediation of clinical reasoning has been conceptualised as requiring explicit metacognitive scaffolding — teaching trainees to monitor their own reasoning processes, recognise schema failures, and apply deliberate corrective strategies — rather than simply increasing exposure to clinical material (Kalet & Chou, 2014; Ramani & Leinster, 2008).

## 6. Discussion

In this paper, we propose a conceptual reframing of a widely observed but imprecisely understood phenomenon in AI-assisted clinical training, that is, the developmental harm produced when medical trainees use AI in ways that are inappropriate to their stage of clinical reasoning development, and the nature of their tasks. The reframing from misuse to misclassification shifts the explanatory locus from behaviour to cognition, from compliance to competency, and from restriction to instruction. Also, it generates specific, testable predictions about the mechanisms of developmental harm as well as the conditions for its prevention.

### 6.1. Theoretical Contributions

First, the introduction of epistemic agency and epistemic responsibility as normative dimensions of the two Developmental Trajectories in the Age of GenAI that SCAN accounts for adds a clinical ethics dimension that, to the best of our knowledge, pure developmental frameworks have not previously incorporated.

Secondly, the concept of misclassification provides the precision that existing framings lack. It specifies what makes a given learner-AI interaction developmentally harmful, how that harm varies across task types, learner stages, and engagement modes, and more importantly, what the appropriate interventions are at each point.

Lastly, we show that, with SCAN, the triad of skill failure, theoretically, distinguishes three directionally distinct pathways that have previously been conflated under the umbrella of AI-related skill erosion, with consequences for intervention design. Furthermore, we identify passive Aid engagement as a particularly insidious and detection-resistant mis-skilling pathway. We argue this requires task re-identification to Non-Negotiable, with human experts serving as epistemic auditors — opening a new line of theoretical and empirical inquiry.

### 6.2. Practical Contributions

We recognize that our framework offers clinical educators a shared, testable vocabulary for evaluating trainee-AI interactions at the task level, supporting concrete redesign across three domains. First, in curriculum design, it identifies, and thus supports to build, metacognitive classification as a skill to be explicitly taught and practiced. It is consistent with the "verify and trust" paradigm proposed by Abdulnour et al. (2025), and with early evidence that Socratic-style AI interfaces sustain greater active engagement than standard AI tools (Wang et al., 2025). Secondly, it restructures supervision to maintain observability of reasoning processes with instruments such as think-aloud protocols in supervision (Tang et al., 2026). Lastly, it develops assessment approaches that can detect which process-level developmental failures (e.g., never-skilling and mis-skilling) before they manifest in output-level performance deficits. We believe this task-level specificity offers the AI-related skill failure that fixed-phase, cohort-wide curricular structures fail to.

### 6.3. Broader Implications

The broader implications extend beyond clinical training to any professional educational domain in which generative AI is being integrated at a pace that outstrips the conceptual frameworks available to evaluate its effects. We believe the misclassification problem is not specific to medical training: it applies wherever AI provides ready substitutes for the very cognitive processes that professional development requires (Risko & Gilbert, 2016). In the medical domain, however, the stakes are the highest: mis-skilled clinical reasoning is not only an educational failure, but also a patient safety concern. The urgency of the reframing proposed here is, therefore, acute.

### 6.4. Limitations

We acknowledge four limitations when implementing our approach. First, the SCAN framework requires metacognitive capacity that novice learners may not yet possess: first-year students asked to classify tasks into SCAN zones may lack the domain-specific knowledge necessary to evaluate what a task genuinely requires — a constraint consistent with evidence that metacognitive accuracy develops in tandem with, not prior to, domain expertise (Kruger & Dunning, 1999; Schraw & Moshman, 1995). We suggest it could be resolved by introducing the framework with high supervisory support from educators before expecting independent application, similar to Phase 1 of Ke et al. (2026)'s three-phase framework. Secondly, implementation also depends on supervisor training and institutional willingness to invest in process-oriented, metacognitive debriefing, which are more demanding than output-focused supervision that currently predominates. Third, SCAN itself remains empirically untested: the framework's central constructs — subzone classification accuracy, engagement quality, and the mis-skilling pathway — await the validation studies outlined in our research agenda (Section 5). Next, never-skilling, as Ke et al. (2026) indicated, is not an established phenomenon. It requires direct, clinical empirical evidence in the future. Lastly, like Ke et al. (2026), we note that this paper reflects the state of medical education and AI at the time of writing — a fast-moving intersection in which specific claims may require revision as evidence accumulates. Until then, the framework's claims should be read as theoretically grounded hypotheses rather than established findings.

## 7. Conclusion

The question of how to integrate generative AI into clinical training without undermining the clinical reasoning that training is designed to produce is one of the most consequential challenges facing medical education today (Boscardin et al., 2024; Triola & Rodman, 2025). The field has correctly identified the issue, and aimed to resolve it at the *structural* level: Quillen College of Medicine has established equitable AI access and policy; Geisel School of Medicine has built a longitudinal AI curriculum; Mass General Brigham has developed health-system governance principles; The AAMC has articulated guiding principles for AI use in medical education. We recognize this question is a cognitive and behavioral one, and thus requires a cognitive and behavioral solution that is mechanistic and falsifiable. What SCAN provides, we believe, is a framework that, for the moment, matters most: the moment before a learner opens an AI tool — when they must decide what kind of engagement is appropriate for the task in front of them, given their current stage of development. It is in this very moment where one's clinical reasoning is formed or deformed; the same moment, we believe, where the field's attention must ultimately arrive.

The paradigm shift from misuse to misclassification is, in the end, a crucial one in what we ask of clinical educators, supervisors, and assessors. Instead of asking for monitoring and restricting, this paradigm shift asks to understand, teach, and evaluate a new *cognitive competency* that is as foundational to clinical practice in the AI era, as the history-taking and physical examination skills that have always defined it. This shift, like any theoretical reframing, must earn its claim through evidence rather than assertion — a claim to be tested, rather than assumed.

**Funding Note.** None declared.

**Conflict of Interest Statement.** None declared.

**Ethical approval**. Reported as not applicable.

**Disclaimers**. None declared.

**Previous presentations.** None declared.

**Data availability.** Reported as not applicable.

**Author contributions (with CRediT details)**:

Conceptualization: Fendi Tsim (FT), Alina Gutoreva (AG)

Project Administration: FT

Visualization: FT

Writing - original draft: FT & AG

Writing - review & editing: FT, AG, Anthony Weiss & Nicole Dubosh